\documentclass[conference]{IEEEtran}
\IEEEoverridecommandlockouts

\usepackage{cite}
\usepackage{amsmath,amssymb,amsfonts}
\usepackage{graphicx}
\usepackage{textcomp}
\usepackage{xcolor}
\usepackage{subcaption}
\usepackage{mathtools}
\usepackage{algorithm}
\usepackage{algpseudocode}
\usepackage{multirow}
\usepackage[hidelinks]{hyperref}

\definecolor{tueScharlaken}{HTML}{C81919}

\def\BibTeX{{\rm B\kern-.05em{\sc i\kern-.025em b}\kern-.08em
    T\kern-.1667em\lower.7ex\hbox{E}\kern-.125emX}}
\begin{document}

\setlength{\textfloatsep}{1mm} 

\title{Hybrid Real- and Complex-Valued Neural Network Architecture for Speech Enhancement\\
\thanks{This work was supported by the Robust AI for SafE (radar) signal processing (RAISE) collaboration framework between Eindhoven University of Technology and NXP Semiconductors, including a Privaat-Publieke Samenwerkingen-toeslag (PPS) supplement from the Dutch Ministry of Economic Affairs and Climate Policy.}
}

\makeatletter
\newcommand{\linebreakand}{%
  \end{@IEEEauthorhalign}
  \hfill\mbox{}\par
  \mbox{}\hfill\begin{@IEEEauthorhalign}
}
\makeatother

\newlength{\extrawidth}
\settowidth{\extrawidth}{\textit{iiiii}}

\newlength{\extrawidthb}
\settowidth{\extrawidthb}{\textit{iiIIIIIiiiiii}}


\author{\IEEEauthorblockN{Luan Vinícius Fiorio}
\IEEEauthorblockA{\textit{Department of Electrical Engineering} \\
\textit{Eindhoven University of Technology}\\
Eindhoven, The Netherlands \\
l.v.fiorio@tue.nl}
\and
\IEEEauthorblockN{Alex Young}
\IEEEauthorblockA{\hspace{\extrawidth}\textit{NXP Semiconductors}\hspace{\extrawidth}\\
Eindhoven, The Netherlands \\
alex.young@nxp.com}
\and
\IEEEauthorblockN{Ronald M. Aarts}
\IEEEauthorblockA{\textit{Department of Electrical Engineering} \\
\textit{Eindhoven University of Technology}\\
Eindhoven, The Netherlands \\
R.M.Aarts@tue.nl}
}

\maketitle

\begin{abstract}
This paper investigates hybrid real- and complex-valued neural networks for monaural speech enhancement. While complex-valued models can process time-frequency representations natively, they often increase computational cost and can be inefficient in small-model regimes. We therefore study a matched-parameter hybrid architecture that combines a real-valued magnitude-mask branch with a complex-valued additive correction branch, coupled via domain conversion functions at the bottleneck. The approach is applied to convolutional denoising autoencoder and convolutional-recurrent network architectures. Averaged over four SNRs, the hybrid models improve intelligibility and quality over the considered counterparts, while substantially reducing the number of operations.
\end{abstract}

\begin{IEEEkeywords}
Neural networks, speech enhancement, complex processing, low-complexity
\end{IEEEkeywords}

\section{Introduction}
\label{sec:intro} 

Speech enhancement aims to improve the quality of noisy speech signals by suppressing or removing unwanted background noise; in the monaural case, this is done from a single-channel recording \cite{Loizou2013Speech}. For this task, time-frequency representations, generally obtained via the short-term Fourier transform (STFT), are widely used as the Fourier transform and its inverse can be computed efficiently \cite{Benesty2011Speech}. 

Some approaches discard the phase component to simplify data handling and reduce complexity \cite{Tan2018Convolutional, Fiorio2024spectral}; however, this can limit intelligibility and perceptual quality since accurate phase reconstruction is crucial for high-quality audio reproduction\cite{sarroff2018complex}. Moreover, independently processing the real and imaginary outputs of a complex-valued STFT can lead to additional distortion and signal degradation \cite{lee2022complex}. Naturally, complex-valued neural networks (NNs) have been applied to speech enhancement as a solution for distorted phase \cite{sarroff2018complex, welker2022interspeech}. While offering richer representational capacity and improved generalisation \cite{trabelsi2018deep}, complex (hereafter `complex') networks can exhibit slower training rates due to automatic differentiation \cite{lee2022complex} and higher model complexity \cite{wu2023rethinking}. 

The scarcely explored combination of real-valued (hereafter `real') and complex processing was previously attempted by averaging results of real and complex models \cite{popa2018deep} or by combining real and complex processing in a two-stage network \cite{du2023ahybrid}, obtaining superior metrics. A related line of work separates magnitude- and phase-oriented processing. For example, PHASEN \cite{yin2020phasen} uses two communicating streams dedicated to amplitude and phase prediction. DeepFilterNet and DeepFilterNet2 combine magnitude-related enhancement with complex filtering mechanisms for efficient real-time processing \cite{schroter2022deepfilternet,schroter2022deepfilternet2}. Our work differs from these approaches in that the separation is explicitly tied to the numerical domain of the network. This makes the model a controlled test case for studying whether hybrid processing can improve parameter-matched model performance and reduce operation count.

We investigate a hybrid alternative to purely real or complex speech-enhancement models. The hybrid framework and domain conversion functions were introduced in our previous work \cite{young2025hybridrealcomplexvaluedneural}; the present paper studies how this concept can be specialised for monaural speech enhancement. The speech-enhancement setting is particularly relevant because magnitude and phase play different perceptual and algorithmic roles. Instead of treating the STFT as a single real tensor or relying exclusively on complex layers, we separate the processing into a real branch that estimates a magnitude mask and a complex branch that estimates an additive complex correction. The main contribution of this paper is therefore a speech-enhancement-specific instantiation and matched-complexity evaluation of this design.

\section{System and Problem Description}

\label{sec:preliminaries}

Let $S(t,f)$, $V(t,f)$, and $Y(t,f)$ denote the complex short-term Fourier transforms (STFTs) of the clean speech $s(t)$, noise $v(t)$, and noisy speech $y(t)$, where $t$ and $f$ are the discrete time and frequency indices respectively, and $Y(t,f) = S(t,f) + V(t,f).$ Moreover, let $M(t,f)$ be a ratio mask \cite{Wang2014OnTraining,Williamson2016complex} which allows us to approximate the clean speech STFT from the noisy speech STFT as $S(t,f) \approx M(t,f) \odot Y(t,f)$ and $\odot$ represents the Hadamard product. In this work, a NN-estimated mask $\hat{M}(t,f)$ is applied to the noisy speech as  $\hat{S}(t,f) = \hat{M}(t,f) \odot Y(t,f)$. For the special case of hybrid networks, a magnitude mask $\hat{M}_{mag}(t,f)$ and a complex correction $\hat{S}_{cc}$ are simultaneously estimated, applied to the noisy STFT as $\hat{S}(t,f) = \hat{M}_{mag}(t,f) \odot Y(t,f) + \hat{S}_{cc}$. For the rest of the manuscript, the indices $(t,f)$ are omitted for simplicity. Note that we consider one complex parameter equivalent to two real parameters when counting, given that a complex parameter is formed by two real parameters $a$ and $b$, as $a + i b$, where $i = \sqrt{-1}$.

\section{Hybrid Neural Network for Speech Enhancement}
\label{sec:hnns}

To investigate the parallel combination of real and complex processing, we propose a generic framework for hybridising a neural network into both real and complex branches. By combining the strengths of both domains, we allow for more targeted processing while reducing model complexity relative to its counterparts, maintaining the same number of parameters. Additionally, we include domain conversion functions, that exchange information between branches. In this work, we focus on autoencoder-based models, which are commonly used for speech enhancement \cite{Fiorio2024spectral} and are straightforward to extend into hybrid architectures. However, the proposed methodology is broadly applicable to a wide range of architectures. Next, we define the design procedure for a hybrid network.

\subsection{Design methodology}
\label{ssec:design}

Let the baseline real model consist of an encoder-decoder pair
$f_{\mathbb{R}}^r:\mathbb{R}^{d_{\text{in}}}\!\to\!\mathbb{R}^{d_z}$ and
$g_{\mathbb{R}}^r:\mathbb{R}^{d_z}\!\to\!\mathbb{R}^{d_{\text{out}}}$, with $N_f$ and $N_g$ (real) parameters, respectively. We construct a
\emph{hybrid} model by decomposing the representation learning and
reconstruction tasks into coupled real and complex branches while keeping
the \emph{realequivalent} parameter count constant. 

The hybrid model is defined by two encoders $f_{\mathbb{R}}^{h}$, $f_{\mathbb{C}}^{h}$ and two decoders $g_{\mathbb{R}}^{h}$, $g_{\mathbb{C}}^{h}$ with explicit cross-domain interactions. We define the real encoder-decoder pair $f_{\mathbb{R}}^{h}$ and $g_{\mathbb{R}}^{h}$ by replicating the baseline architecture, but rescaling the layer widths such that the number of parameters become $N_f/2$ and $N_g/2$ respectively. This branch preserves the inductive bias of the original model and reserves capacity for a complex branch. In parallel, we instantiate a complex encoder-decoder pair $f_{\mathbb{C}}^{h}$ and $g_{\mathbb{C}}^{h}$ with the same layer arrangement and activation placement as the real branch, but using complex parameters. We modify layer widths such that the numbers of complex trainable parameters are $N_f/4$ and $N_g/4$, respectively, since one complex parameter is counted as two real parameters. Consequently, the hybrid model preserves the total realequivalent number of parameters. If a recurrent bottleneck is present, we account for those layers as part of the encoder.

The equal realequivalent split is chosen as a neutral design point rather than as an optimised hyperparameter. It avoids assigning a larger capacity to either numerical domain a priori and makes the comparison with the real and complex baselines interpretable. Other splits may be beneficial depending on the task and model size. Evaluating such alternatives is left for future ablation studies. The present goal is to determine whether a simple, balanced hybridisation already provides consistent gains under a fixed parameter budget. 

Since the hybrid network processes both real and complex tensors, we introduce domain conversion functions: deterministic, algebraic mappings that do not introduce trainable parameters. For latent branch interaction, we use Cartesian domain conversion functions, which map between real and complex representations while preserving information and shape compatibility:
\begin{subequations}
    \begin{equation}
        \phi_{\mathbb{C}\rightarrow
        \mathbb{R}}:\mathbb{C}^{m}\to\mathbb{R}^{2m},
        \quad 
        \phi_{\mathbb{C}} \rightarrow \mathbb{R} (\mathbf{z}) \triangleq
        \begin{bmatrix}
            \Re(\mathbf{z})\\ 
            \Im(\mathbf{z})
        \end{bmatrix},
    \end{equation}
    \begin{equation}
        \phi_{\mathbb{R}\rightarrow \mathbb{C}}:\mathbb{R}^{2m}\to\mathbb{C}^{m},\quad
        \phi_{\mathbb{R}\rightarrow \mathbb{C}}\!\left(\!\begin{bmatrix}\mathbf{a}\\ \mathbf{b}\end{bmatrix}\!\right)
        \triangleq \mathbf{a}+j\mathbf{b}. 
    \end{equation}    
\end{subequations}
At the hybrid network input, $X$ is complex. Because the real branch requires a real input, we apply a magnitude domain conversion function (`Mag' in Figure~\ref{fig:hybrid_models}). This removes phase content from the STFT data and allows the real branch to use a magnitude-only representation for denoising. This operation is lossy and changes the shape of the data, halving the number of channels. We refer to \cite{young2025hybridrealcomplexvaluedneural} for an extensive exploration of domain conversion functions and their influence in hybrid networks.

Furthermore, let $\mathbf{h}_{\mathbb{R}} \triangleq f_{\mathbb{R}}^{h}(\mathbf{x}_{\mathbb{R}})\in\mathbb{R}^{d_{\mathbb{R}}}$, and $\mathbf{h}_{\mathbb{C}} \triangleq f_{\mathbb{C}}^{h}(\mathbf{x}_{\mathbb{C}})\in\mathbb{C}^{d_{\mathbb{C}}}$. We couple the branches via symmetric concatenation with conversion, as $\tilde{\mathbf{h}}_{\mathbb{R}} \triangleq \mathrm{cat}\!\left(\mathbf{h}_{\mathbb{R}},\, \phi_{\mathbb{C}\rightarrow \mathbb{R}}(\mathbf{h}_{\mathbb{C}})\right)$, $\tilde{\mathbf{h}}_{\mathbb{C}} \triangleq \mathrm{cat}\!\left(\mathbf{h}_{\mathbb{C}},\, \phi_{\mathbb{R}\rightarrow \mathbb{C}}(\mathbf{h}_{\mathbb{R}}^{\uparrow})\right)$, where $\mathrm{cat}(\cdot,\cdot)$ denotes concatenation along the feature dimension, and $\mathbf{h}_{\mathbb{R}}^{\uparrow}\in\mathbb{R}^{2d_{\mathbb{C}}}$ denotes a shape-adjusted version of $\mathbf{h}_{\mathbb{R}}$ so that $\phi_{\mathbb{R}\rightarrow \mathbb{C}}$ is well-defined.

The conversion $\phi_{\mathbb{R}\rightarrow\mathbb{C}}$ should not be interpreted as recovering the STFT phase from the real branch. Instead, it reinterprets pairs of learned real latent features as the real and imaginary coordinates of an abstract complex latent representation. The purpose of the conversion is to provide the complex decoder with structured information learned by the real branch, allowing subsequent complex layers to rotate and combine this information in the complex domain.

\begin{figure}[!t]
\centering
    \begin{subfigure}{.47\textwidth}
        \centering
        \includegraphics[width=0.925\textwidth]{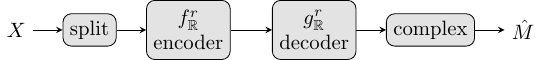}
        \caption{Real model}
        \label{fig:real_models}
    \end{subfigure} \\ \vspace{5mm}
    \begin{subfigure}{.47\textwidth}
        \centering
        \includegraphics[width=0.58\textwidth]{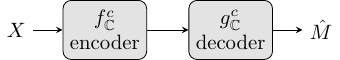}
        \caption{Complex model}
        \label{fig:complex_models}
        \vspace{2.5mm}
    \end{subfigure}
    \begin{subfigure}{.47\textwidth}
        \centering
        \includegraphics[width=0.99\textwidth]{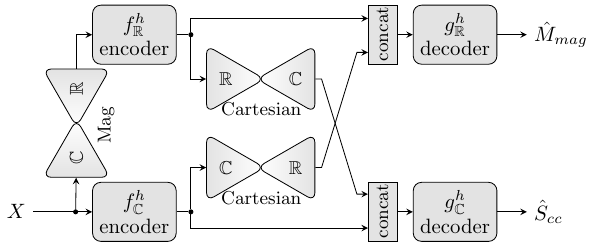}
        \caption{Hybrid model}
        \label{fig:hybrid_models}
    \end{subfigure}
\caption{Real, complex, and hybrid models. Blocks with an hourglass shape are domain conversion functions, Cartesian conversions are used at the bottleneck, while `Mag' denotes a magnitude conversion applied at the input. $\mathrm{concat}$ means concatenation, $\mathrm{split}$ is $\mathrm{split}(X) = \mathrm{concat}(\mathcal{R}(X), \mathcal{I}(X))$, and $\mathrm{complex}$ is the opposite of $\mathrm{split}$.}
\label{fig:architectures_denoising}
\end{figure}
    
The decoders then operate on the fused representations $\hat{\mathbf{y}}_{\mathbb{R}} = g_{\mathbb{R}}^{h}(\tilde{\mathbf{h}}_{\mathbb{R}})$, $\hat{\mathbf{y}}_{\mathbb{C}} = g_{\mathbb{C}}^{h}(\tilde{\mathbf{h}}_{\mathbb{C}})$. Because concatenation increases the input dimensionality to each decoder, the first layer of $g_{\mathbb{R}}^{h}$ and $g_{\mathbb{C}}^{h}$ is modified (e.g., by reducing its output width), such that the overall decoder budgets remain $N_g/2$ and $N_g/4$, respectively, i.e., each branch retains half of the original realequivalent decoder capacity. This yields a coupled hybrid architecture in which each decoder uses both its native-domain features and the domain-translated features learned by the complementary branch. We also derive a complex baseline (superscript $c$) from the real model, with an encoder $f_\mathbb{C}^c$ and decoder $g_\mathbb{C}^c$ with $N_f/2$ and $N_g/2$ real parameters, respectively. All three cases are shown in Figure~\ref{fig:architectures_denoising}. 

\begin{table*}[!t]
    \centering
    \caption{CDAE and CRN layers' channel size. Subscripts $\mathbb{R}$ and $\mathbb{C}$ refer to real and complex branches, respectively.}
    \begin{tabular}{c c c | c c c c c}
        \hline
        \textbf{Model} & \textbf{Conv2D} & \textbf{Conv2D.T} & \textbf{Model} & \textbf{Conv2D} & \textbf{GRU} & \textbf{Linear} & \textbf{Conv2D.T} \\
        \hline
        rCDAE & [16, 32, 64, 128] & [64, 32, 16, 1] & rCRN & [16, 32, 64, 128] & [96, 96] & 1536 & [64, 32, 16, 1]\\
        
        cCDAE & [16, 18, 44, 96] & [44, 18, 16, 1]  & cCRN & [16, 22, 44, 64] & [110, 112] & 512 & [44, 22, 16, 1] \\
        
        hCDAE$_\mathbb{R}$ & [16, 18, 44, 96] & [22, 14, 8, 1]  & hCRN$_\mathbb{R}$ & [22, 24, 44, 64] & [110, 110] & 512 & [24, 16, 8, 1] \\
        
        hCDAE$_\mathbb{C}$ & [8, 16, 32, 64] & [20, 14, 8, 1] & hCRN$_\mathbb{C}$ & [8, 16, 32, 48] & [76, 76] & 384 & [22, 14, 8, 1]\\
        \hline
    \end{tabular}
    \label{tab:nn_sizes}
    \vspace{-2mm}
\end{table*}

For speech enhancement, we combine the hybrid model's real output $\hat{M}_{mag}$ with the complex output $\hat{S}_{cc}$ as $\hat{S} = \hat{M}_{mag} \odot Y + \hat{S}_{cc}$, resulting in magnitude filtering and complex residual correction. We implement the complex correction using addition such that the magnitude mask handles most of the denoising, similar to state-of-the-art methods \cite{schroter2022deepfilternet, schroter2022deepfilternet2}. The complex term provides fine-grained compensation of phase and magnitude components, avoiding possible distortion created from multiplicative filtering. The complex layers and activation functions are detailed next.




\subsection{Complex Layers and Activations}

We define complex layers based on \cite{trabelsi2018deep}. A complex layer $l_c(\cdot)$ is implemented via complex multiplication using two separate real layers of the same type $l_\mathbb{R}^1(\cdot)$ and $l_\mathbb{R}^2(\cdot)$ as
\begin{equation}
\label{eq:layers}
    l_c(Z) = l_\mathbb{R}^1(\mathcal{R}(Z)) - l_\mathbb{R}^2(\mathcal{I}(Z)) + i (l_\mathbb{R}^1(\mathcal{I}(Z)) + l_\mathbb{R}^2(\mathcal{R}(Z))),
\end{equation}
for a complex input tensor $Z$. When formulating recurrent layers, the use of \eqref{eq:layers} follows \cite{hu2020dccrn} and corresponds to representing a complex multiplication using real matrices \cite{trabelsi2018deep}. For consistency, we also use \eqref{eq:layers} to define complex recurrent layers.

In terms of activation functions, we follow the approach in \cite{young2025hybridrealcomplexvaluedneural}, where constrained polynomials are used. We consider the complex ReLU (cReLU) function, implemented as 
\begin{equation}
    \mathrm{cReLU}(Z) = \frac{Z}{2} + \frac{Z^2}{2(|Z| + 0.01)}.
\end{equation}
Additionally, we employ the complex Tanh (cTanh) function, implemented as 
\begin{equation}
    \mathrm{cTanh}(Z) = \frac{Z}{\sqrt{|Z|^2 + 1}}.
\end{equation}
The forms of $\mathrm{cReLU}$ and $\mathrm{cTanh}$ are particularly relevant for stability because the denominators constrain the growth of the numerators. This ensures that, in the limit as $|Z| \rightarrow \infty$, the functions exhibit at most linear growth. Note that, if $Z$ is substituted by a real tensor, $\mathrm{cReLU}$ and $\mathrm{cTanh}$ reduce to real approximations of ReLU and Tanh.

\section{Numerical Experiments}
\label{sec:experiments}

We compare two architectures, the convolutional denoising autoencoder (CDAE) and the convolutional-recurrent network (CRN), in their real, complex, and hybrid forms, based on their real implementation from \cite{Fiorio2024spectral}. Data, architectures, results, and analysis are detailed next.

\subsection{Data}

The raw audio from LibriTTS (speech reference) and TAU2019 (urban acoustic noise) are converted to a sampling frequency of 16 kHz. Each 10 s noise segment is combined with 10 s of concatenated, randomly drawn clean-speech files at a uniformly sampled SNR between -5 and 20 dB. Raised cosine fade-in and fade-out are applied to all files, with durations sampled uniformly between 0.2 and 0.3 s. The time-frequency domain signals are obtained via STFT using a 256-sample Hann window, an FFT of the same length and 50\% overlap. Magnitude warping is applied to the STFT output $Y$ using a log magnitude function $W = max(-80, 20 \log_{10}(|Y| + \epsilon))$, clipped below -80 dB, then $X = (W/80+1)Y/(|Y| + \epsilon)$, compressing the dB range of -80 to 0 dB to a linear scale of 0 to 1, $\epsilon = 10^{-8}$.

For training, the LibriTTS train-clean-100 subset (100 h) and the entire development set of the TAU2019 dataset (40 h) are used to generate 100 h of augmented audio. Evaluation is performed over 20 h of augmented data with the LibriTTS test-clean subset (8.6 h) and the TAU2019 evaluation set (20 h). Note that this evaluation protocol is intended for controlled comparison between real, complex, and hybrid variants under identical training and test conditions.

\begin{table*}[!t]
    \centering
    \caption{Performance metrics obtained by the models over the test set for different SNR levels. Best values highlighted in bold.}
    \resizebox{\textwidth}{!}{%
    \begin{tabular}{c c c c c c c c c c c c c c}
        \hline
        
        & \multicolumn{3}{c}{-5 dB SNR} & \multicolumn{3}{c}{0 dB SNR} & \multicolumn{3}{c}{10 dB SNR} & \multicolumn{3}{c}{20 dB SNR} \\
        
        \textbf{Method} & \textbf{STOI} & \textbf{PESQ} & \textbf{SI-SDR} & \textbf{STOI} & \textbf{PESQ} & \textbf{SI-SDR} & \textbf{STOI} & \textbf{PESQ} & \textbf{SI-SDR} & \textbf{STOI} & \textbf{PESQ} & \textbf{SI-SDR} \\
        
        \hline        
        
        \textcolor{darkgray}{None (noisy)} & \textcolor{darkgray}{0.714} & \textcolor{darkgray}{1.441} & \textcolor{darkgray}{-3.199} & \textcolor{darkgray}{0.811} & \textcolor{darkgray}{1.667} & \textcolor{darkgray}{1.801} & \textcolor{darkgray}{0.939} & \textcolor{darkgray}{2.405} & \textcolor{darkgray}{11.801} & \textcolor{darkgray}{0.986} & \textcolor{darkgray}{3.356} & \textcolor{darkgray}{21.801} \\

        \hline
                
        rCDAE & 0.714 & 1.514 & 2.048  & 0.819 & 1.794 & 6.867  & 0.944 & 2.621 & 15.397  & 0.986 & 3.553 & 22.094 \\
        
        cCDAE & 0.734 & 1.580 & 3.689  & 0.836 & 1.902 & 8.460  & 0.951 & 2.798 & 16.794  & 0.987 & \textbf{3.708} & 23.463  \\
        
        hCDAE & \textbf{0.743} & \textbf{1.599} & \textbf{4.404}  & \textbf{0.844} & \textbf{1.949} & \textbf{9.127}  & \textbf{0.953} & \textbf{2.855} & \textbf{17.288}  & \textbf{0.988} & 3.703 & \textbf{24.119}  \\
        
        \hline
                
        rCRN & 0.734 & 1.608 & 4.219  & 0.838 & 1.960 & 8.783  & 0.952 & 2.872 & 17.003  & 0.988 & 3.679 & 24.400  \\
        
        cCRN & 0.793 & 1.853 & 7.411  & 0.881 & 2.291 & 11.644  & 0.965 & 3.249 & 19.149  & \textbf{0.991} & 3.948 & 26.258  \\
        
        hCRN & \textbf{0.801} & \textbf{1.862} & \textbf{7.762}  & \textbf{0.887} & \textbf{2.314} & \textbf{11.977}  & \textbf{0.967} & \textbf{3.281} & \textbf{19.353}  & \textbf{0.991} & \textbf{3.957} & \textbf{26.286} \\
        \hline    
    \end{tabular}}
    \label{tab:results}
    \vspace{-2mm}
\end{table*}

\subsection{Architectures}

Our baseline is the real CDAE (rCDAE) from \cite{Fiorio2024spectral}. The rCDAE takes as input the concatenation of real and imaginary parts of the normalized noisy-speech STFT $X$, and estimates the real and imaginary parts of a complex mask \cite{williamson2016crm}, which is recombined into a complex tensor. The convolutions operate only along the frequency dimension, which reduces complexity and preserves time-invariant operation. We also consider the real convolutional recurrent network (rCRN) from \cite{Fiorio2024spectral}, reducing the layer sizes to lower the total number of parameters. The complex and hybrid models (cCDAE, hCDAE, cCRN, and hCRN) are obtained as proposed in Section~\ref{ssec:design}. For all networks, ReLU (or cReLU if complex) is used after each convolutional (Conv2D) or transposed convolutional (Conv2D.T) layers, except for the last encoder and decoder layers. The last encoder layer contains a Tanh (or cTanh if complex), and the last decoder layer contains no activation functions for all models except for the real branch of the hCDAE and hCRN, which uses a Sigmoid, as they estimate a magnitude mask. The layer sizes of all networks are shown in Table~\ref{tab:nn_sizes}, where each number in a square bracket represents a layer's output size, ordered from left to right.

The output strategy differs across model families because it follows the intended role of each architecture. In this paper, we focus on the comparison between conventional single-branch mask estimation and the proposed hybrid decomposition under matched parameter counts. We therefore interpret the gains as resulting from the combined design of hybrid branch specialisation, bottleneck coupling, and residual complex correction, rather than from the coupling mechanism alone.

\subsection{Performance}

All models are trained for 100 epochs each, using Adam with weight decay of $10^{-4}$ and an initial learning rate of $10^{-3}$, decaying exponentially to $10^{-4}$ by the last epoch. We use the (negative) scale-invariant signal-to-distortion ratio (SI-SDR) for the loss function, as it has previously shown to be effective for speech enhancement \cite{roux2018sdrhalfbakeddone,kolbaek2020sisdr}. We analyse the performance in terms of SI-SDR, short-time objective intelligibility (STOI) \cite{taal2010stoi}, and perceptual evaluation of speech quality (PESQ) \cite{rix2001pesq}. 

The performance of all CDAE- and CRN-based models for different SNR conditions is shown in Table~\ref{tab:results}. For both architectures, the real variants consistently underperform their complex and hybrid models. In particular, for rCDAE at -5 dB, the STOI value is identical to the noisy input, indicating that the rCDAE improves PESQ and SI-SDR but does not improve objective intelligibility. This behaviour is consistent with the limited capacity of a small real model operating on split real and imaginary STFT components. The complex baseline improves over the real models in our matched-parameter setting. This observation differs from the general conclusion of \cite{wu2023rethinking}, which reports that complex operations may be inefficient or even detrimental for small monaural speech-enhancement models. We attribute this difference to the experimental setting and architecture: our real baselines are compact frequency-convolutional models with a restricted operation pattern, whereas \cite{wu2023rethinking} evaluates different architectures and complex building blocks. Therefore, our results should not be interpreted as contradicting the broader finding that purely complex networks are often computationally inefficient. Instead, they show that, for the considered CDAE and CRN baselines, native complex processing is beneficial but still not the most efficient option once MACs are considered.

The hCDAE and hCRN outperform their real and complex counterparts in nearly all reported conditions. The largest absolute gains over the real baselines occur in SI-SDR, suggesting that the hybrid decomposition is particularly helpful for reconstructing the enhanced waveform with lower distortion. The gains over the complex baselines are smaller but consistent, indicating that the hybrid model is not merely recovering the advantage of complex processing, but also benefits from separating magnitude filtering from complex residual correction.

The additive form is motivated by residual estimation. A multiplicative complex mask must jointly encode magnitude suppression, phase rotation, and possible reconstruction corrections in a single factor. In contrast, the proposed decomposition assigns most attenuation to the bounded real magnitude mask and reserves the complex branch for residual compensation. This may be easier to optimise, especially for small models, because the complex branch does not need to reproduce the entire clean-speech STFT. We emphasize, however, that multiplicative complex correction and different branch-capacity allocations are natural ablation directions and are not claimed to be inferior in general.

From another perspective, the gains of the hybrid models may also be related to their output structure. This is consistent with state-of-the-art architectures \cite{schroter2022deepfilternet, schroter2022deepfilternet2}, which separate magnitude processing from phase filtering, achieving better speech-related metrics than that of models which process magnitude and phase in a unified branch. This output structure is more closely related to residual enhancement than to conventional complex-mask estimation. The magnitude mask provides a stable denoising path, while the additive complex term compensates for errors that are difficult to represent through a bounded or multiplicative mask alone. However, because the current experiments do not isolate the output structure from the hybrid coupling, the relative contribution of each component remains an open question.

\begin{table}[!t]
    \centering
    \caption{MACs for each of the considered models. [$\mathbb{R}$] and [$\mathbb{C}$] represent, respectively, MACs calculated from the real and complex branches of the neural network model. The lowest total MAC values are highlighted in bold.}
    \begin{tabular}{c c c c c}
        \hline
        \textbf{Model} & \textbf{Param.} & \textbf{MACs} [$\mathbb{R}$] &  \textbf{MACs} [$\mathbb{C}$] & \textbf{MACs} \\
        \hline
        rCDAE & 173.3 k & 4.72 G & 0 & 4.72 G \\
        cCDAE & 171.5 k & 0 & 4.54 G & 4.54 G \\
        hCDAE & 172.2 k & 1.08 G & 2.23 G & \textbf{3.31 G} \\ 
        \hline
        rCRN & 816 k & 6.88 G & 0 & 6.88 G \\
        cCRN & 815 k & 0 & 8.04 G & 8.04 G \\
        hCRN & 816 k & 1.80 G & 3.71 G & \textbf{5.51 G} \\
        \hline
    \end{tabular}
    \label{tab:macs}
    \vspace{0mm}
\end{table}

\subsection{Complexity Analysis}
\label{ssec:complexity}
The number of parameters of each model and their computational complexity in terms of multiply-accumulate operations (MACs) are shown in Table~\ref{tab:macs}. A complex multiplication requires multiple real operations, therefore, we count MACs in terms of the underlying real operations used to implement each complex layer. Note that, in Figure~\ref{fig:architectures_denoising}, only transposed convolution layers are considered part of the decoder. Convolutional, recurrent, and linear layers are grouped with the encoder for counting purposes.

For the CDAE models, rCDAE and cCDAE have comparable MAC counts, whereas hCDAE requires only 3.31 G MACs, reducing 29.9\% relative to rCDAE and 27.1\% relative to cCDAE. For the CRN models, hCRN requires 5.51 G MACs, corresponding to reductions of 19.9\% relative to rCRN and 31.5\% relative to cCRN. The cCRN is more expensive than rCRN despite having a matched parameter count, mainly because recurrent complex operations expand into several real operations. These results also demonstrate that parameter count alone is insufficient for assessing efficiency. In the proposed hybrid models, splitting the network into smaller real and complex branches substantially reduces dimensionality, thereby lowering the total operation count while preserving the representational benefit of complex processing.

\section{Conclusion}

We investigated hybrid real and complex neural networks for monaural speech enhancement by constructing matched-parameter complex and hybrid counterparts of CDAE and CRN baselines. The proposed hybrid models combine real magnitude masking with additive complex correction, while exchanging information between branches using Cartesian domain conversion functions at the bottleneck. In a controlled proof of concept experiment, the hybrid models achieved the best STOI, PESQ, and SI-SDR values in almost all SNR conditions. Compared with purely complex models, the gains were modest but consistent, while MAC counts were reduced by 27.1\% for hCDAE and 31.5\% for hCRN. These results indicate that the proposed hybrid realcomplex architecture can offer a favourable trade-off between enhancement performance and computational complexity.

\bibliographystyle{IEEEtran}
\bibliography{refs}

@ARTICLE{Williamson2016complex,
  author={Williamson, Donald S. and Wang, Yuxuan and Wang, DeLiang},
  journal={IEEE/ACM Transactions on Audio, Speech, and Language Processing}, 
  title={Complex Ratio Masking for Monaural Speech Separation}, 
  year={2016},
  volume={24},
  number={3},
  pages={483-492},
  doi={10.1109/TASLP.2015.2512042}}

@ARTICLE{Wang2014OnTraining,
  author={Wang, Yuxuan and Narayanan, Arun and Wang, DeLiang},
  journal={IEEE/ACM Transactions on Audio, Speech, and Language Processing}, 
  title={On Training Targets for Supervised Speech Separation}, 
  year={2014},
  volume={22},
  number={12},
  pages={1849-1858},
  doi={10.1109/TASLP.2014.2352935}}

@misc{young2025hybridrealcomplexvaluedneural,
      title={Hybrid Real- and Complex-valued Neural Network Architecture}, 
      author={Alex Young and Luan Vinícius Fiorio and Bo Yang and Boris Karanov and Wim van Houtum and Ronald M. Aarts},
      year={2025},
      eprint={2504.03497},
      archivePrefix={arXiv},
      primaryClass={cs.LG},
      url={https://arxiv.org/abs/2504.03497}, 
}

@INPROCEEDINGS{schroter2022deepfilternet,
  author={Schroter, Hendrik and Escalante-B, Alberto N. and Rosenkranz, Tobias and Maier, Andreas},
  booktitle={2022 IEEE International Conference on Acoustics, Speech and Signal Processing (ICASSP)}, 
  title={Deepfilternet: A Low Complexity Speech Enhancement Framework for Full-Band Audio Based On Deep Filtering}, 
  year={2022},
  volume={},
  number={},
  pages={7407-7411},
  doi={10.1109/ICASSP43922.2022.9747055}}

@INPROCEEDINGS{schroter2022deepfilternet2,
  author={Schröter, H. and Maier, A. and Escalante-B, A.N. and Rosenkranz, T.},
  booktitle={2022 International Workshop on Acoustic Signal Enhancement (IWAENC)}, 
  title={Deepfilternet2: Towards Real-Time Speech Enhancement on Embedded Devices for Full-Band Audio}, 
  year={2022},
  volume={},
  number={},
  pages={1-5},
  doi={10.1109/IWAENC53105.2022.9914782}}

@inproceedings{trabelsi2018deep,
  author       = {Chiheb Trabelsi and
                  Olexa Bilaniuk and
                  Ying Zhang and
                  Dmitriy Serdyuk and
                  Sandeep Subramanian and
                  Jo{\~{a}}o Felipe Santos and
                  Soroush Mehri and
                  Negar Rostamzadeh and
                  Yoshua Bengio and
                  Christopher J. Pal},
  title        = {Deep Complex Networks},
  booktitle    = {6th International Conference on Learning Representations, {ICLR} 2018,
                  Vancouver, BC, Canada, April 30 - May 3, 2018, Conference Track Proceedings},
  year         = {2018},
}

@ARTICLE{Fiorio2024spectral,
  author={Fiorio, Luan Vinícius and Karanov, Boris and Defraene, Bruno and David, Johan and Widdershoven, Frans and Van Houtum, Wim and Aarts, Ronald M.},
  journal={IEEE Access}, 
  title={Spectral Masking With Explicit Time-Context Windowing for Neural Network-Based Monaural Speech Enhancement}, 
  year={2024},
  volume={12},
  number={},
  pages={154843-154852},
  doi={10.1109/ACCESS.2024.3483443}}

@INPROCEEDINGS{williamson2016crm,
  author={Williamson, Donald S. and Wang, Yuxuan and Wang, DeLiang},
  booktitle={2016 IEEE International Conference on Acoustics, Speech and Signal Processing (ICASSP)}, 
  title={Complex ratio masking for joint enhancement of magnitude and phase}, 
  year={2016},
  volume={},
  number={},
  pages={5220-5224},
  doi={10.1109/ICASSP.2016.7472673}}

@article{roux2018sdrhalfbakeddone,
      title={{SDR - half-baked or well done?}}, 
      author={Jonathan Le Roux and Scott Wisdom and Hakan Erdogan and John R. Hershey},
      year={2018},
      journal={arXiv preprint 1811.02508},
      primaryClass={cs.SD} 
}

@article{kolbaek2020sisdr,
author = {Kolb\ae{}k, Morten and Tan, Zheng-Hua and Jensen, S\o{}ren Holdt and Jensen, Jesper},
title = {On Loss Functions for Supervised Monaural Time-Domain Speech Enhancement},
year = {2020},
issue_date = {2020},
publisher = {IEEE Press},
volume = {28},
issn = {2329-9290},
url = {https://doi.org/10.1109/TASLP.2020.2968738},
doi = {10.1109/TASLP.2020.2968738},
journal = {IEEE/ACM Trans. Audio, Speech and Lang. Proc.},
month = feb,
pages = {825–838},
numpages = {14}
}

@INPROCEEDINGS{taal2010stoi,
  author={Taal, Cees H. and Hendriks, Richard C. and Heusdens, Richard and Jensen, Jesper},
  booktitle={2010 IEEE International Conference on Acoustics, Speech and Signal Processing}, 
  title={A short-time objective intelligibility measure for time-frequency weighted noisy speech}, 
  year={2010},
  volume={},
  number={},
  pages={4214-4217},
  doi={10.1109/ICASSP.2010.5495701}}

@INPROCEEDINGS{rix2001pesq,
  author={Rix, A.W. and Beerends, J.G. and Hollier, M.P. and Hekstra, A.P.},
  booktitle={2001 IEEE International Conference on Acoustics, Speech, and Signal Processing. Proceedings (Cat. No.01CH37221)}, 
  title={Perceptual evaluation of speech quality (PESQ)-a new method for speech quality assessment of telephone networks and codecs}, 
  year={2001},
  volume={2},
  number={},
  pages={749-752 vol.2},
  doi={10.1109/ICASSP.2001.941023}}

@misc{wu2023rethinking,
      title={Rethinking complex-valued deep neural networks for monaural speech enhancement}, 
      author={Haibin Wu and Ke Tan and Buye Xu and Anurag Kumar and Daniel Wong},
      year={2023},
      eprint={2301.04320},
      archivePrefix={arXiv},
      primaryClass={cs.SD}
}

@phdthesis{sarroff2018complex,
  title        = {{C}omplex {N}eural {N}etworks for {A}udio --- digitalcommons.dartmouth.edu},
  author       = {Andy M. Sarroff},
  year         = 2018,
  month        = {May},
  school       = {Dartmouth College},
  type         = {PhD thesis}
}

@ARTICLE{lee2022complex,
  author={Lee, ChiYan and Hasegawa, Hideyuki and Gao, Shangce},
  journal={IEEE/CAA Journal of Automatica Sinica}, 
  title={Complex-Valued Neural Networks: A Comprehensive Survey}, 
  year={2022},
  volume={9},
  number={8},
  pages={1406-1426},
}

@book{Loizou2013Speech,
author = {Loizou, Philipos C.},
title = {Speech Enhancement: Theory and Practice},
year = {2013},
isbn = {1466504218},
publisher = {CRC Press, Inc.},
address = {USA},
edition = {2nd},
}

@book{Benesty2011Speech,
author = {Benesty, Jacob and Chen, Jingdong and Habets, Emanul A.P.},
title = {Speech Enhancement in the STFT Domain},
year = {2011},
isbn = {3642232493},
publisher = {Springer Publishing Company, Incorporated},
edition = {1st},
}

@inproceedings{Tan2018Convolutional,
  author={Ke Tan and DeLiang Wang},
  title={{A Convolutional Recurrent Neural Network for Real-Time Speech Enhancement}},
  year=2018,
  booktitle={Proc. Interspeech 2018},
  pages={3229--3233},
  doi={10.21437/Interspeech.2018-1405}
}

@inproceedings{welker2022interspeech,
  author={Simon Welker and Julius Richter and Timo Gerkmann},
  title={{Speech Enhancement with Score-Based Generative Models in the Complex STFT Domain}},
  year=2022,
  booktitle={Proc. Interspeech 2022},
  pages={2928--2932},
}

@INPROCEEDINGS{popa2018deep,
  author={Popa, Călin-Adrian},
  booktitle={2018 International Joint Conference on Neural Networks (IJCNN)}, 
  title={Deep Hybrid Real-Complex-Valued Convolutional Neural Networks for Image Classification}, 
  year={2018},
  volume={},
  number={},
  pages={1-6},
}

@article{du2023ahybrid,
title = {A hybrid complex-valued neural network framework with applications to electroencephalogram (EEG)},
journal = {Biomedical Signal Processing and Control},
volume = {85},
pages = {104862},
year = {2023},
issn = {1746-8094},
author = {Hang Du and Rebecca Pillai Riddell and Xiaogang Wang},
}

@inproceedings{hu2020dccrn,
  author    = {Hu, Yanxin and Liu, Yun and Lv, Shubo and Xing, Mengtao and Zhang, Shimin and Fu, Yihui and Wu, Jian and Zhang, Bihong and Xie, Lei},
  title     = {{DCCRN}: Deep Complex Convolution Recurrent Network for Phase-Aware Speech Enhancement},
  booktitle = {Proc. Interspeech},
  pages     = {2472--2476},
  year      = {2020},
  doi       = {10.21437/Interspeech.2020-2537}
}

@inproceedings{yin2020phasen,
  author    = {Yin, Dacheng and Luo, Chong and Xiong, Zhiwei and Zeng, Wenjun},
  title     = {{PHASEN}: A Phase-and-Harmonics-Aware Speech Enhancement Network},
  booktitle = {Proceedings of the AAAI Conference on Artificial Intelligence},
  volume    = {34},
  number    = {05},
  pages     = {9458--9465},
  year      = {2020}
}

\end{document}